\documentclass[11pt,letterpaper]{article}

\usepackage[margin=1in]{geometry}
\usepackage{microtype}
\usepackage{graphicx}
\usepackage{subcaption}
\usepackage{booktabs}
\usepackage{amsmath}
\usepackage{amssymb}
\usepackage{mathtools}
\usepackage{xcolor}
\usepackage[colorlinks=true,linkcolor=blue!60!black,citecolor=blue!60!black,urlcolor=blue!60!black]{hyperref}
\usepackage{xurl}  
\usepackage[numbers,sort&compress]{natbib}
\usepackage{setspace}
\usepackage{parskip}

\usepackage[T1]{fontenc}
\usepackage{lmodern}

\usepackage{float}
\usepackage[labelfont=bf,font=small]{caption}

\newcommand{\R}{\mathbb{R}}
\newcommand{\bz}{\mathbf{z}}
\newcommand{\bh}{\mathbf{h}}
\newcommand{\bW}{\mathbf{W}}
\DeclareMathOperator*{\softmax}{softmax}

\title{\textbf{UMA-Inverse: Ligand-Conditioned Protein Inverse Folding
       with a Distogram-Supervised Dense Pair Encoder}}

\author{William Sobolewski\\
        \texttt{wcsobo7@gmail.com}}

\date{July 8, 2026}

\begin{document}
\maketitle

\begin{abstract}
Designing protein sequences that bind specific ligands benefits from an inverse-folding model conditioned
on full ligand geometry. We present \textbf{UMA-Inverse},
which replaces the sparse graph backbone of LigandMPNN with a dense pair-representation encoder: a
six-block PairMixer (triangle multiplication, no triangle self-attention or sequence track) refines
all residue--residue and residue--ligand atom pairs, supervised by an auxiliary \emph{distogram}
objective, and an autoregressive decoder attends over ligand atoms through a learned, position-specific
readout of the pair tensor. The model is compact ($\sim$3.3\,M parameters). On the LigandMPNN test
splits it reaches 56.1\%/55.1\%/35.3\% interface recovery
(small-molecule/metal/nucleotide). It trails LigandMPNN, but by less than the published numbers
suggest: re-run under our identical protocol, LigandMPNN scores $59.8$/$64.4$/$53.3$ (vs.\ published
$63.3$/$77.5$/$50.5$). In a
pocket-fixed setting the redesigns are confidently folded and ligand-binding-competent under Boltz-2
cofolding, again modestly behind LigandMPNN. Its distinctive property is representational: the dense
encoder propagates ligand identity to residues far beyond the interface, where LigandMPNN's signal
decays. We offer UMA-Inverse as a compact baseline for ligand-conditioned inverse folding
that trails LigandMPNN in accuracy, together with a characterization of how a dense all-pairs encoder
distributes ligand information.
\end{abstract}

\section{Introduction}

Protein sequence design, specifying the amino acid sequence that adopts a desired fold, is a
central task in computational protein engineering. Since the introduction of
ProteinMPNN~\citep{dauparas2022proteinmpnn}, autoregressive graph neural network (GNN) based inverse
folding has become the dominant approach: given fixed backbone coordinates, a message-passing
network conditions per-residue predictions on local structural context before decoding amino acids
in a causal order. LigandMPNN~\citep{dauparas2025ligandmpnn} extends this framework to
ligand-conditioned design by inserting ligand heavy atoms as extra graph nodes, and is the most
widely adopted and experimentally validated model for interface sequence recovery across metal,
nucleotide, and small-molecule ligand classes. It remains the standard baseline for the task,
though several more recent and substantially larger models report higher recovery on the same
splits~\citep{yi2025adflip}.

Despite its success, the sparse $k$-nearest-neighbor graph in LigandMPNN places an inherent
constraint on long-range information flow. Distant residues communicate only through chains of
local message-passing steps, which may limit the model's ability to encode allosteric and
shell-by-shell geometric relationships. Separately, the AlphaFold2
Evoformer~\citep{jumper2021alphafold} demonstrated that maintaining an explicit $N{\times}N$ pair
tensor, updated by triangle multiplication and transition MLP layers, provides powerful geometric
reasoning by capturing three-body relationships among all residue triplets. AlphaFold3 showed this
architecture extends naturally to small-molecule and metal cofactors~\citep{abramson2024alphafold3}.
A natural question is whether dense pair representations can improve ligand-conditioned inverse
folding.

We present \textbf{UMA-Inverse}\footnote{The ``UMA'' name is personal and unrelated to Meta FAIR's UMA
(Universal Models for Atoms).}, a ligand- and nucleic-acid-conditioned inverse-folding model that
substitutes the GNN encoder with a six-block PairMixer-inspired encoder: a stripped-down Pairformer that
retains triangle multiplication but omits triangle self-attention, following the observation that
triangle multiplication alone recovers nearly all of the geometric inductive bias at substantially
lower computational cost~\citep{ouyangzhang2025pairmixer}. Our encoder is a from-scratch
reimplementation of the PairMixer block design rather than the released module. Two design choices distinguish UMA-Inverse from a bare
dense-encoder baseline, each targeting a specific failure mode of pair-tensor inverse folding:
(i)~an \textbf{auxiliary distogram objective} that supervises the encoder to make its pair tensor
explicitly structure-predictive; and (ii)~a \textbf{learned ligand-attention decoder} in which each
residue reads ligand context through a position-specific softmax over the pair tensor rather than a
uniform mean-pool. We additionally route DNA/RNA chains into the ligand atom pool to probe nucleotide
partners; this conditioning fails to improve on a ligand-agnostic baseline, for an architectural reason we analyze in
\S\ref{sec:discussion}. The resulting model is compact
(${\sim}3.3$\,M parameters) and produces sequences that fold with high Boltz-2 cofolding confidence.

Our main contributions are:
\begin{enumerate}
  \item A PairMixer-based inverse-folding architecture with explicit residue--residue and
        residue--ligand pair representations, supervised by an auxiliary distogram head that makes
        the encoder's geometry explicit (\S\ref{sec:methods}).
  \item A learned ligand-attention decoder that conditions each residue's prediction on
        position-specific ligand context (\S\ref{sec:arch}).
  \item A systematic, reproducible evaluation covering teacher-forced recovery, autoregressive
        interface recovery across all three ligand classes, constrained pocket design, Boltz-2
        cofolding, and distance-shell ligand-signal analysis (\S\ref{sec:results}), on which
        LigandMPNN remains more accurate than UMA-Inverse on recovery and cofolded pose.
  \item An empirical characterization of the resulting representation: the dense all-pairs encoder
        propagates ligand identity to residues far beyond the binding interface (several-fold to
        ${\sim}30\times$ the distal signal of LigandMPNN's local graph), a property we measure across
        all three ligand classes (\S\ref{sec:results_distal}).
\end{enumerate}

\section{Background and Related Work}
\label{sec:background}

\paragraph{Autoregressive inverse folding.}
ProteinMPNN~\citep{dauparas2022proteinmpnn} decodes amino acids in a random order, conditioning each
prediction on an already-decoded subset via causal attention over graph edges. LigandMPNN extends
this by treating ligand atoms as additional nodes with element-type features and adds a
separate distance-encoding from backbone atoms to ligand heavy atoms. Both models use sparse
$k$-NN message-passing with a fixed $k{=}48$ neighbors. GVP-GNN~\citep{jing2021gvp} is a
related approach that encodes geometric vector quantities directly; ESM-IF~\citep{hsu2022esmif}
couples such a geometric encoder with a large sequence model trained on predicted structures, and
PiFold~\citep{gao2023pifold} is a one-shot (non-autoregressive) graph design model. None of these
condition on ligand context.

\paragraph{Dense pair representations.}
AlphaFold2~\citep{jumper2021alphafold} introduced the Evoformer, which jointly updates a sequence
representation $\bh_i$ and a pair representation $\bz_{ij}$ via triangle multiplication, triangle
self-attention, and outer-product mean. AlphaFold3 simplified this into the Pairformer. Recent work
shows triangle self-attention is the computationally dominant and empirically redundant component:
models retaining only triangle multiplication and transition MLP layers recover essentially the same
accuracy at substantially reduced cost~\citep{ouyangzhang2025pairmixer}. We adopt this simplified architecture and,
distinctively, supervise the pair tensor directly with a distogram loss.
The closest architectural precedent is CarbonDesign~\citep{ren2024carbondesign}, which applies an
AlphaFold2-style pair representation to (protein-only) inverse folding, encoding the backbone into
single and pair representations through triangular edge-update layers before decoding with an
amortized Markov random field. UMA-Inverse departs from it on two axes: it admits ligand and
nucleic-acid atoms into the same pair tensor to make the design ligand-conditioned, and it retains
only triangle multiplication, omitting the triangle self-attention that CarbonDesign's edge updates
use.

\paragraph{Auxiliary structural supervision.}
Distogram prediction, classifying binned inter-residue distances from a pair
representation, was a core training signal in AlphaFold2~\citep{jumper2021alphafold} and AlphaFold3.
We repurpose it as an \emph{auxiliary} objective for inverse folding: the primary loss remains
sequence cross-entropy, but a lightweight distogram head encourages the encoder to retain explicit
geometric structure in its pair tensor, which the decoder then reads.

\paragraph{Ligand-conditioned structure prediction and design.}
AlphaFold3~\citep{abramson2024alphafold3} and Boltz-2~\citep{passaro2025boltz2} solve the forward
problem (structure prediction given sequence and ligand) using diffusion over atomic coordinates.
RFdiffusion~\citep{watson2023rfdiffusion} tackles backbone generation but requires a separate inverse-folding
step. RFdiffusion3~\citep{butcher2025rfdiffusion3} has begun folding sequence design into the generative
process itself (its all-atom diffusion co-generates side-chain atoms, implicitly specifying residue
identities alongside the backbone), but the directly diffused sequences remain less structure-consistent
than dedicated inverse folding, and the authors still fit sequences with ProteinMPNN/LigandMPNN for
their benchmarks. Dedicated inverse folding thus remains the standard sequence-design step even inside
all-atom generative pipelines. UMA-Inverse occupies that niche: given a fixed protein--ligand backbone,
it predicts the most probable complementary sequence.

\paragraph{Ligand- and nucleic-acid-conditioned inverse folding.}
LigandMPNN is the most widely adopted and experimentally validated model for this task, but as of
mid-2026 it is no longer the sole accuracy frontier on its own benchmark. ADFLIP~\citep{yi2025adflip},
an all-atom discrete flow-matching model, reports higher interface recovery than LigandMPNN on the
identical small-molecule, nucleotide, and metal test splits used here, albeit with a substantially
larger, iterative generative model. Caliby and its ligand-conditioned Potts
variant~\citep{shuai2025caliby} augment LigandMPNN's ligand encoder and report large gains on
AlphaFold3-evaluated ligand-placement metrics: a self-consistency rather than native-recovery
criterion, and a distinction we return to in \S\ref{sec:discussion}. CARBonAra~\citep{krapp2024carbonara}
is a parameter-free geometric transformer that conditions on ligands, ions, and nucleic acids, and
NA-MPNN~\citep{kubaney2025nampnn} unifies protein, DNA, and RNA design in a single biopolymer graph,
directly targeting the nucleic-acid setting our routing addresses. Independent benchmarks of ligand-conditioned inverse folding are also beginning to appear:
\citet{wei2025ifbenchmark} rank five models (including ProteinMPNN, LigandMPNN, and CARBonAra) across
25{,}716 curated protein--ligand complexes from BioLiP. We benchmark against LigandMPNN as
the field's standard baseline; these are recent and concurrent work we cite for positioning but do
not re-evaluate. Against this landscape UMA-Inverse's distinguishing property is not accuracy but
compactness (a $\sim$3.3\,M-parameter, single-pass model), together with the dense-encoder
representation we characterize below.

\section{Methods}
\label{sec:methods}

\subsection{Dataset}
\label{sec:data}

We use the same training, validation, and test splits as LigandMPNN~\citep{dauparas2025ligandmpnn},
derived from the PDB with cutoffs at 30\% sequence identity between splits. PDB files are fetched
from RCSB and parsed by a vendorized, extended version of the LigandMPNN parser
(\texttt{src/data/pdb\_parser.py}). We do not crop residues to a pocket around the ligand: the full
protein chain is retained as designable context. A separate \emph{size} limit applies for batch
efficiency (structures whose residue-plus-ligand node count exceeds 384 are cropped to 384 nodes), and
up to \textbf{50} ligand heavy atoms nearest the protein centroid are kept per structure (raised from 25
to accommodate larger ligands and nucleic-acid context). The 384-node cap bounds total structure size
for batching and is distinct from pocket cropping; it defaults to the same value at inference and is
configurable for larger targets.

\paragraph{Nucleic-acid routing.}
DNA and RNA chains appear in PDB files as \texttt{ATOM} records (not \texttt{HETATM}) with residue
names \texttt{DA/DC/DG/DT} (DNA) or \texttt{A/C/G/U} (RNA). A parse that collects ligands only from
\texttt{HETATM} records therefore misses them as ligands, while an amino-acid parse skips them as
non-standard residues; without special handling they fall out of both pools. We extend the parser to
route these nucleotide atoms into the ligand atom pool, which enables evaluation on the nucleotide test
split (\S\ref{sec:results_iface}). Ligand atoms enter as a single global
set capped at 50 (the nearest to the protein centroid), which covers a compact ligand but only a small
fragment of a nucleic acid. As \S\ref{sec:results_iface} shows, this is insufficient, and in fact
counterproductive, for nucleotide partners; the cause and a fix are discussed in \S\ref{sec:discussion}.

\paragraph{Augmentation.}
Backbone and ligand coordinates receive Gaussian noise ($\sigma{=}0.1$\,{\AA}) during training,
simulating coordinate uncertainty and regularizing against memorization of exact PDB geometry. With
probability 0.03 per designable residue, the residue's native sidechain heavy atoms are appended to
the ligand context, exposing the model to real sidechain geometry during decoding.

\subsection{Model Architecture}
\label{sec:arch}

\begin{figure}[t]
  \centering
  \includegraphics[width=\linewidth]{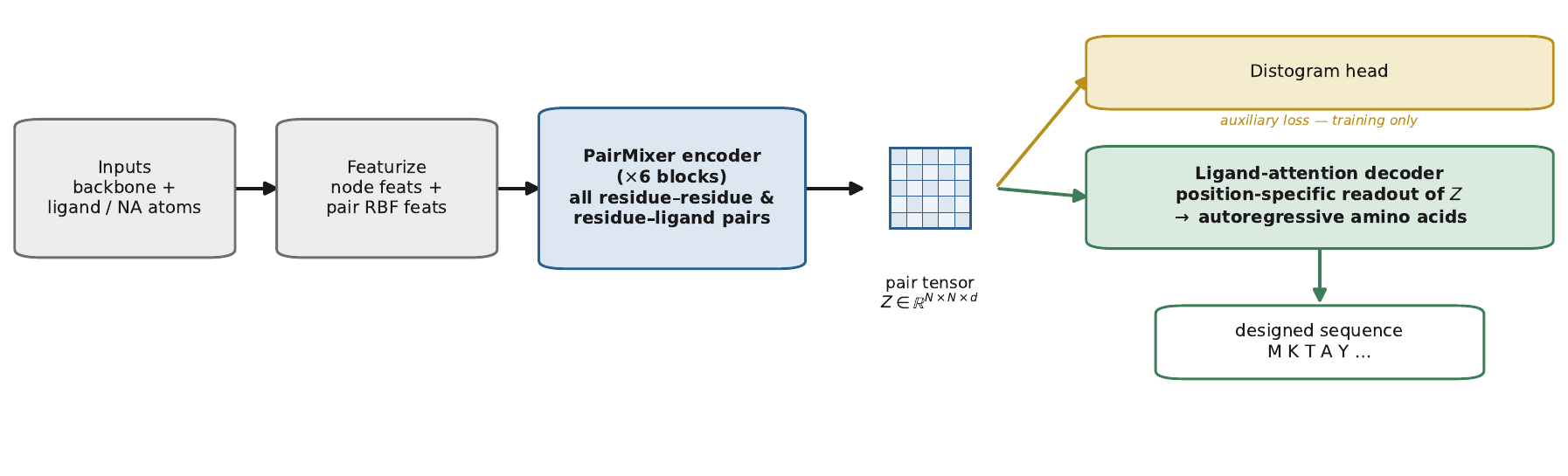}
  \caption{\textbf{UMA-Inverse architecture.} Backbone and ligand/nucleic-acid atoms are featurized
    into node and pairwise (RBF distance) features; a six-block PairMixer encoder refines the pair
    tensor $Z$ over all residue--residue and residue--ligand atom pairs. Each block uses triangle
    multiplication and a pair transition and omits triangle self-attention \citep{ouyangzhang2025pairmixer}.
    An auxiliary distogram head supervises $Z$ during training, and an autoregressive decoder reads
    ligand context from $Z$ through a learned, position-specific attention to emit the sequence.}
  \label{fig:architecture}
\end{figure}

UMA-Inverse maps a protein--ligand backbone to per-residue amino acid logits via four stages
(Fig.~\ref{fig:architecture}): featurization, pair-tensor initialization, PairMixer encoding, and
autoregressive decoding.

\paragraph{Featurization.}
Each residue $i$ receives a 6-dimensional feature vector of the sine and cosine of the backbone
dihedral angles $(\phi_i, \psi_i, \omega_i)$. Ligand (and nucleotide) atoms are featurized by the
LigandMPNN atomic featurizer: a concatenation of atom-type, group, and period one-hot vectors,
projected to $d{=}128$ by a learned linear layer. For residue--residue pairs we compute 25 pairwise
backbone distances (the \texttt{backbone\_full\_25} protocol); for residue--ligand pairs we compute
five backbone-atom-to-ligand-atom distances and encode frame-relative sine/cosine angles from each
residue's local frame to each ligand atom.

\paragraph{Pair-tensor initialization.}
Node features are projected to $d{=}128$ to form initial node representations
$\bh_i \in \R^{128}$. The pair tensor $\bz_{ij} \in \R^{128}$ is initialized as
$\bz_{ij} = W_L\,\bh_i + W_R\,\bh_j + W_{\text{rbf}}\,\phi_{\text{rbf}}(d_{ij}) + \mathbf{r}_{ij}$,
where $\phi_{\text{rbf}}(d_{ij}) \in \R^{32}$ is a 32-Gaussian RBF embedding of the inter-node
distance over $[0,24]$\,{\AA}, and $\mathbf{r}_{ij}$ is a learned relative-position embedding of the
clamped sequence offset $\mathrm{clip}(i-j,\,\pm 32)$, with a dedicated index for any pair involving a
ligand atom.

\paragraph{PairMixer encoder.}
The pair tensor is refined by 6 PairMixer blocks. Each block applies, in sequence:
(i)~incoming triangle multiplication
$\bz_{ij} \leftarrow \bz_{ij} + \text{LN}(\sum_k g(\bz_{ki}) \odot g(\bz_{kj}))$;
(ii)~outgoing triangle multiplication
$\bz_{ij} \leftarrow \bz_{ij} + \text{LN}(\sum_k g(\bz_{ik}) \odot g(\bz_{jk}))$;
(iii)~a two-layer transition MLP (expansion 4) applied per pair. The gate
$g(\cdot) = \text{Linear}(\cdot) \odot \sigma(\text{Linear}(\cdot))$ follows the AlphaFold2
convention. Each triangle-multiplication update additionally applies an AlphaFold2-style output gate
and an output projection, omitted from the expressions above for brevity; no triangle self-attention
is used. We reimplement the PairMixer block
of~\citet{ouyangzhang2025pairmixer} from scratch, using a standard GELU transition MLP in place of
their gated (SwiGLU) variant and omitting their row/column structured dropout; the triangle
multiplication and gating follow the original formulation. After encoding, a learned
attention-weighted pool over each residue's row of $\bz$ (masked softmax over valid columns) is
projected and added back to the node embedding.

\paragraph{Auxiliary distogram head.}
\label{sec:distogram}
A lightweight head $\text{Linear}(d_{\text{pair}}, 38)$ predicts, for each residue pair $(i,j)$, a
distribution over 38 distance bins spanning $[3.15, 50.75]$\,{\AA} (1.25\,{\AA} wide), targeting the
binned virtual-C$\beta$--C$\beta$ distance (C$\beta$ derived from N/C$\alpha$/C). The model is trained
with the combined objective
\begin{equation}
  \mathcal{L} = \mathcal{L}_{\text{CE}} + \lambda\,\mathcal{L}_{\text{distogram}},
  \qquad \lambda = 0.2,
\end{equation}
where $\mathcal{L}_{\text{CE}}$ is the per-residue sequence cross-entropy and
$\mathcal{L}_{\text{distogram}}$ is the mean cross-entropy of the binned-distance classification over
valid residue pairs (both residues present, $|i-j|\geq 2$). The head is \emph{training-only}: at inference the model returns sequence logits
and the head is discarded (it is ignored by the inference checkpoint loader). Its role is to force the
encoder's pair tensor to remain explicitly structure-predictive, which the ligand-attention decoder
then exploits.

\paragraph{Autoregressive decoder with learned ligand attention.}
Sequences are decoded in a random order during training; non-designed residues are placed earliest so
they provide context. The autoregressive context for residue $i$ is a softmax-weighted aggregation
over previously decoded residues, biased by the residue--residue sub-block of $\bz$
(analogous to ProteinMPNN). The \emph{ligand} context is the key departure from a mean-pool
baseline: rather than averaging the residue--ligand sub-block $\bz_{i,\ell}$ uniformly, each residue
$i$ reads a position-specific, attention-weighted summary,
\begin{equation}
  \alpha_{i\ell} = \softmax_\ell\!\Big(\tfrac{\bz_{i\ell}\,\bW_a}{\sqrt{d_{\text{pair}}}}\Big),
  \qquad
  \mathbf{s}_i = \sum_{\ell} \alpha_{i\ell}\,\bz_{i\ell},
\end{equation}
with $\bW_a = \text{Linear}(d_{\text{pair}}, 1)$ (128 added parameters). The weighted sum is taken over
the raw residue--ligand pair rows; the ligand-context channel fed to the decoder is a learned
projection of $\mathbf{s}_i$ concatenated with a mean-pooled residue--residue context,
$\bh_i^{(\text{lig})} = W_c\,[\,\mathrm{pool}_j\,\bz_{ij}\,\Vert\,\mathbf{s}_i\,]$. This lets interface residues
attend sharply to their coordinating ligand atoms while distal residues integrate the field more
diffusely, rather than every residue receiving the same average. The design was motivated by an
outcome-level KL analysis of the mean-pool baseline, which diverged ${\sim}20{\times}$ from the native
distribution at $>$25\,{\AA} despite encoding strong near-ligand signal. The node representation, the
autoregressive context, and the ligand context enter the decoder as separate concatenated channels,
$\text{logits}_i = \text{MLP}([\,\bh_i \,\Vert\, \bh_i^{(\text{ar})} \,\Vert\, \bh_i^{(\text{lig})}\,]) \in \R^{21}$.

The model has approximately 3.3\,M parameters, comparable to LigandMPNN (${\sim}2.5$\,M); the
distogram head and ligand-attention readout add ${\approx}5$\,k parameters combined.

\subsection{Training}
\label{sec:training}

\begin{figure}[t]
  \centering
  \includegraphics[width=\linewidth]{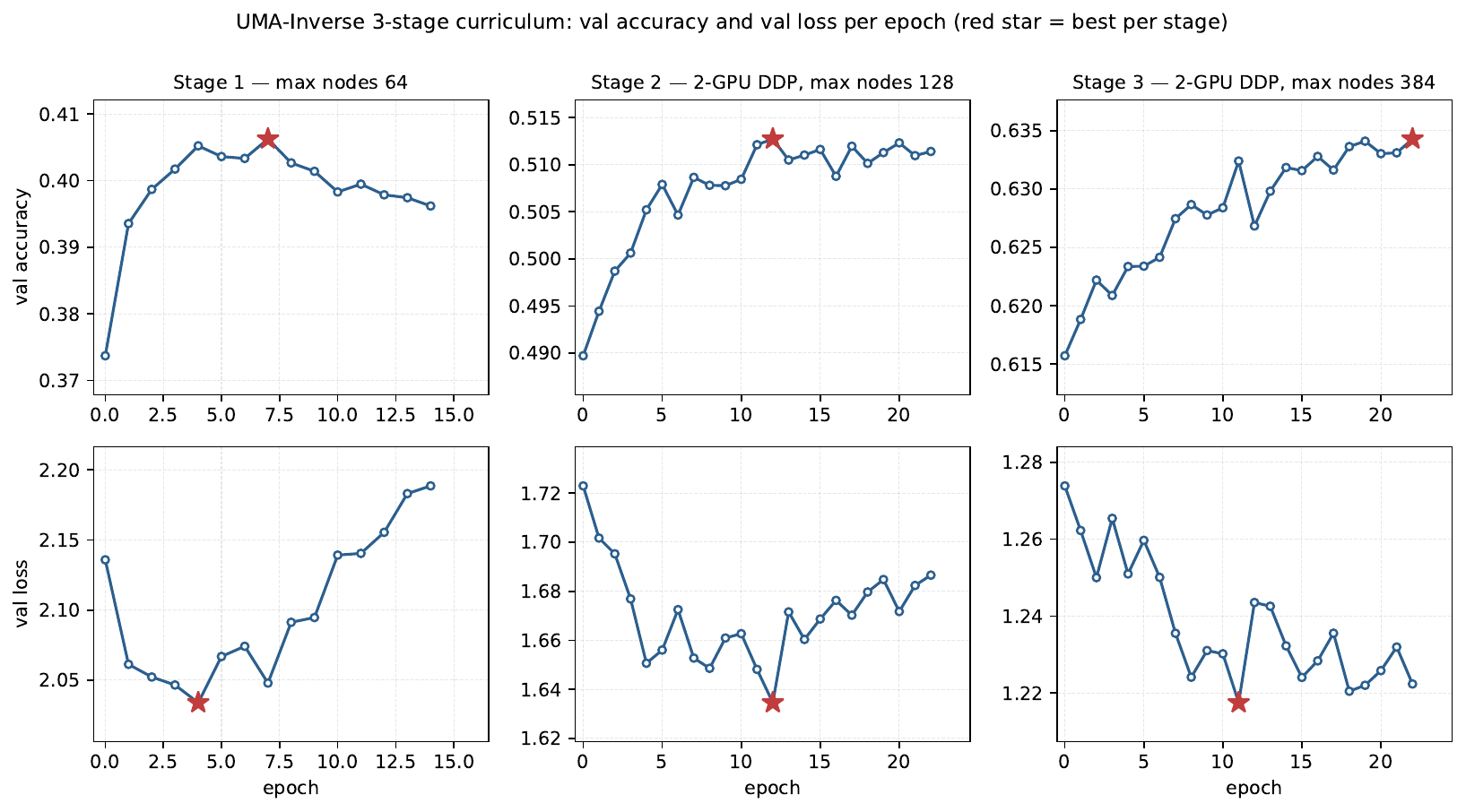}
  \caption{\textbf{Three-stage training curves.} Validation accuracy and loss across the curriculum;
    the auxiliary distogram top-1 accuracy is overlaid for Stage 3. The best checkpoint (red star)
    is at epoch 11.}
  \label{fig:training}
\end{figure}

We train with a three-stage curriculum of increasing structural complexity, progressively raising
the per-example crop size ($64\to128\to384$ nodes) following the staged training schedule
of~\citet{ouyangzhang2025pairmixer}. We optimize with AdamW
\citep{loshchilov2019adamw} (lr $3{\times}10^{-4}$, weight decay $10^{-2}$), linear warmup over
2{,}000 steps, and cosine decay thereafter; gradients clipped at 1.0; bfloat16-mixed precision.
Stage 1 ($\leq$64 nodes) and Stage 2 ($\leq$128 nodes) build up structural context; Stage 3
($\leq$384 nodes, $M{=}50$ ligand atoms, distogram $\lambda{=}0.2$) is full-resolution training on
2$\times$A100 GPUs with effective batch 16 (the final stage taking roughly one week). The best
checkpoint by validation loss is at epoch
11 (val loss 1.22, val accuracy 63.2\%); training is plateaued by this
point, with the auxiliary distogram top-1 accuracy reaching 93.1\%
(Table~\ref{tab:training_stages}, Fig.~\ref{fig:training}). We select by validation loss rather than
accuracy: a mild late-training calibration drift (validation cross-entropy rising while accuracy
plateaus) makes loss the more reliable checkpoint criterion.

\subsection{Inference}
\label{sec:inference}

At inference the autoregressive context is zeroed (as in ProteinMPNN/LigandMPNN), making the decoder
a one-shot predictor; the distogram head is discarded. Sequences are sampled by multinomial sampling
at temperature $T{=}0.1$ unless stated. Pocket-fixed design locks all but the residues within
8\,{\AA} of the ligand and provides the native identity of fixed residues as context.

\subsection{Evaluation protocols}
\label{sec:eval}

\paragraph{Teacher-forced sequence recovery.}
One forward pass per structure using the native sequence as context; recovery is the fraction of
positions where the argmax matches the native residue, excluding unknown (X).

\paragraph{Interface sequence recovery.}
Directly following the LigandMPNN protocol~\citep{dauparas2025ligandmpnn}: ten autoregressive samples per
structure at $T{=}0.1$ with random decoding order and seed 0, recovery restricted to interface
residues (sidechain heavy atom within 5\,{\AA} of any non-protein heavy atom), per-PDB median over
the 10 samples, headline = mean of per-PDB medians. We report all three ligand classes
(small molecule, metal, nucleotide).

\paragraph{Pocket-fixed design and cofolding.}
For a curated set of small-molecule and metal structures, 20 sequences per structure are generated
with the binding pocket ($\leq$8\,{\AA}) fixed to native and the remainder designed; recovery is
reported for the designed (non-pocket) residues. Each designed sequence is then cofolded with Boltz-2
\citep{passaro2025boltz2} (best-of-5 confidence, ipTM, ligand ipTM).

\paragraph{Distogram and distal-signal diagnostics.}
We report the auxiliary head's top-1 binned-distance accuracy on held-out structures, and the native
log-probability elevation as a function of residue-to-ligand distance (5\,{\AA} shells) to quantify
how far ligand identity propagates.

\section{Results}
\label{sec:results}

\subsection{Training convergence}
\label{sec:results_training}

The model reaches validation accuracy 63.2\% at its best checkpoint, and the auxiliary distogram
objective trains to 93.1\% top-1 binned-distance accuracy, indicating the encoder's pair tensor is
strongly structure-predictive. This validation accuracy is the per-residue figure logged during
training on our split (which includes nucleic-acid complexes routed into ligand context); it is not
measured on the same metric or data as the recovery numbers reported for LigandMPNN and so is not
directly comparable. The controlled head-to-head comparison is the matched-protocol interface
recovery of \S\ref{sec:results_iface}.

\subsection{Teacher-forced sequence recovery}
\label{sec:results_tf}

On the full validation set, UMA-Inverse achieves 66.1\% per-PDB mean sequence recovery
(pooled 70.6\%) with perplexity 2.57 and expected calibration error (ECE)
0.0078. Per-amino-acid recovery is reported in Table~\ref{tab:peraacov}; as expected,
backbone-determined residues (Gly, Pro) recover most reliably. Zeroing the ligand features at
evaluation lowers mean recovery by $1.2$\,pp ($66.1\%\!\to\!64.8\%$) and native-sequence
log-likelihood by $0.083$\,nats/residue. Averaged over the whole protein this effect is small (most
residues lie far from the ligand and are largely backbone-determined), but it concentrates near the
binding site, as the distance-shell analysis (\S\ref{sec:results_distal}) makes explicit.

\subsection{Interface sequence recovery}
\label{sec:results_iface}

\begin{figure}[t]
  \centering
  \begin{subfigure}[b]{0.85\linewidth}
    \includegraphics[width=\linewidth]{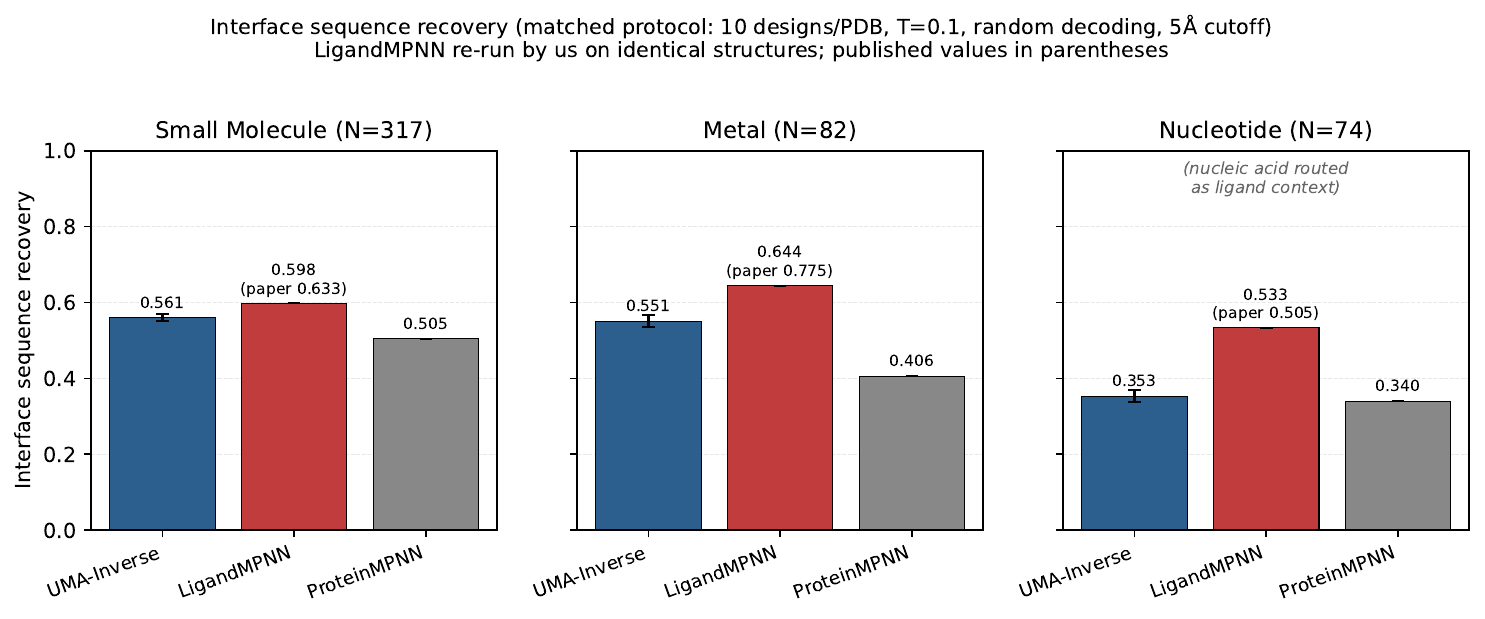}
    \caption{Interface recovery by ligand class and model.}
    \label{fig:iface_bars}
  \end{subfigure}

  \vspace{0.6em}
  \begin{subfigure}[b]{0.85\linewidth}
    \includegraphics[width=\linewidth]{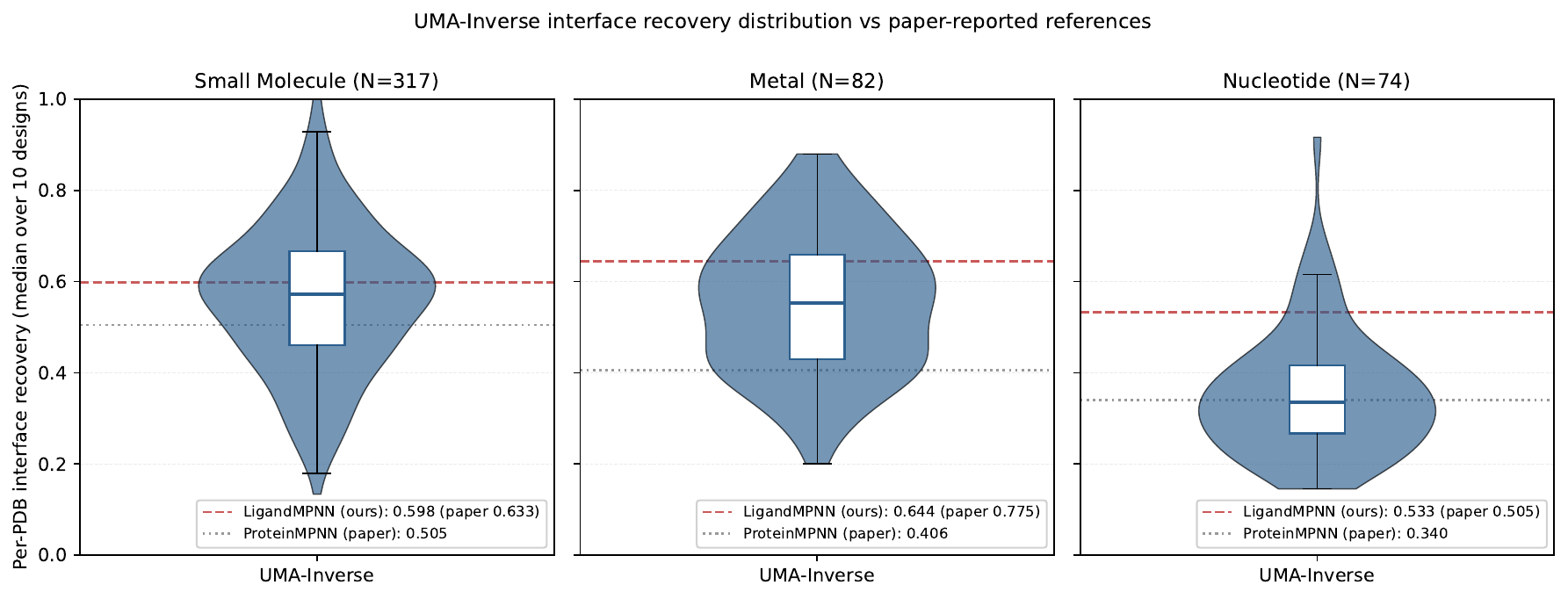}
    \caption{Per-PDB recovery distribution.}
    \label{fig:iface_violins}
  \end{subfigure}
  \caption{\textbf{Interface sequence recovery on the LigandMPNN test splits.}
    (a) Mean-of-per-PDB-medians for UMA-Inverse vs.\ our matched-protocol LigandMPNN run (LigandMPNN's
    published value in parentheses) and the published ProteinMPNN lower bound. (b) Per-PDB recovery
    distributions, with our matched LigandMPNN (dashed) and published ProteinMPNN (dotted) as
    references.}
  \label{fig:iface}
\end{figure}

We evaluate on all three LigandMPNN test splits (Table~\ref{tab:iface}). Because published baselines
depend on protocol details, we also re-run LigandMPNN ourselves on the identical structures under our
exact protocol (same $5$\,{\AA} interface mask, $10$ samples, mean-of-per-PDB-medians) and treat that
matched run as the controlled comparison. Under it, UMA-Inverse trails LigandMPNN by $3.7$\,pp on
small molecule ($56.1\%$ vs.\ $59.8\%$), $9.3$\,pp on metal ($55.1\%$ vs.\ $64.4\%$), and $18.0$\,pp on
nucleotide ($35.3\%$ vs.\ $53.3\%$). Pairing per PDB on the structures scored by both models, the gap
is significant on every split: paired UMA$-$LigandMPNN differences of $-4.1$/$-9.0$/$-17.7$\,pp
(95\% bootstrap CI $[-5.8,-2.5]$/$[-11.7,-6.2]$/$[-21.6,-13.8]$; Wilcoxon signed-rank
$p<10^{-5}$ for all three; $N{=}300$/$72$/$67$). Our matched LigandMPNN numbers are
\emph{lower} than its published values ($0.633$/$0.775$/$0.505$), most strikingly on metal ($0.644$
vs.\ $0.775$), so the gap to UMA-Inverse is markedly smaller under a controlled comparison than the
published numbers imply. We report both, with published values in
parentheses. ProteinMPNN (no ligand conditioning, published)
is a lower bound ($0.505$/$0.406$/$0.340$). On the nucleotide split UMA-Inverse only
matches the ligand-agnostic ProteinMPNN ($35.3\%$ vs.\ $34.0\%$) while trailing LigandMPNN by
$18.0$\,pp. This is the opposite of the small-molecule and metal splits, where conditioning clearly
helps (UMA-Inverse exceeds ProteinMPNN by $5.6$ and $14.5$\,pp): its nucleic-acid conditioning
extracts essentially none of the signal LigandMPNN obtains from the same atoms ($+16.5$\,pp over
ProteinMPNN, published values). We trace this failure to how UMA-Inverse's
dense encoder admits ligand atoms, and discuss it in \S\ref{sec:discussion}.

\begin{table}[h]
\centering
\caption{Interface sequence recovery on the LigandMPNN test splits. Protocol: 10 autoregressive
  samples per PDB, $T{=}0.1$, random order, 5\,{\AA} sidechain--nonprotein cutoff; mean-of-per-PDB-medians.
  \textbf{LigandMPNN is re-run by us under this identical protocol on the same structures} (its
  published value from \citet{dauparas2025ligandmpnn} in parentheses); LigandMPNN (this work) is scored on
  $300$/$72$/$67$ of the PDBs after excluding cases where its parsed native sequence disagreed with
  ours. ProteinMPNN is the published value.}
\label{tab:iface}
\begin{tabular}{lcccc}
\toprule
Split & N & UMA-Inverse & LigandMPNN & ProteinMPNN \\
               &     &             & ours (paper) & (paper) \\
\midrule
Small molecule & 317 & 0.561 & 0.598 (0.633) & 0.505 \\
Metal          & 82  & 0.551 & 0.644 (0.775) & 0.406 \\
Nucleotide     & 74  & 0.353 & 0.533 (0.505) & 0.340 \\
\bottomrule
\end{tabular}
\end{table}

\subsection{Pocket-fixed design and cofolding}
\label{sec:results_pocket}

\begin{figure}[t]
  \centering
  \begin{subfigure}[b]{0.48\linewidth}
    \includegraphics[width=\linewidth]{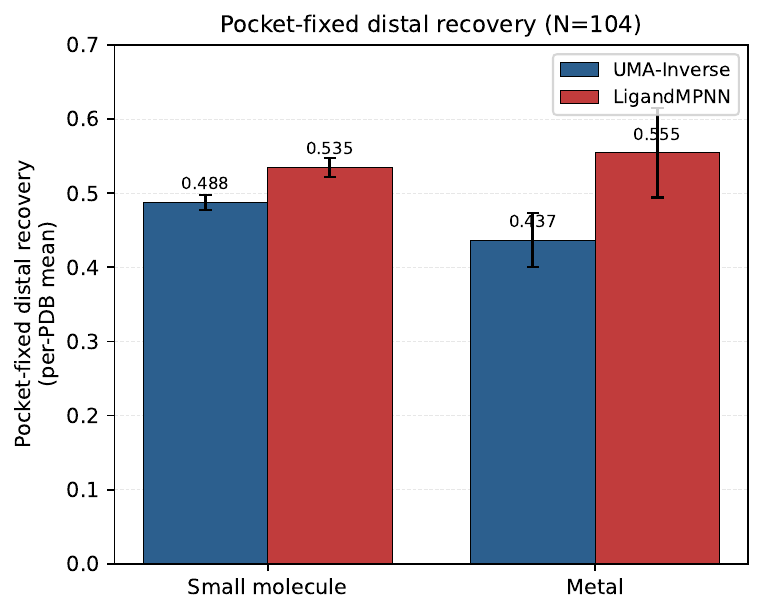}
    \caption{Pocket-fixed distal recovery by ligand class.}
  \end{subfigure}
  \hfill
  \begin{subfigure}[b]{0.48\linewidth}
    \includegraphics[width=\linewidth]{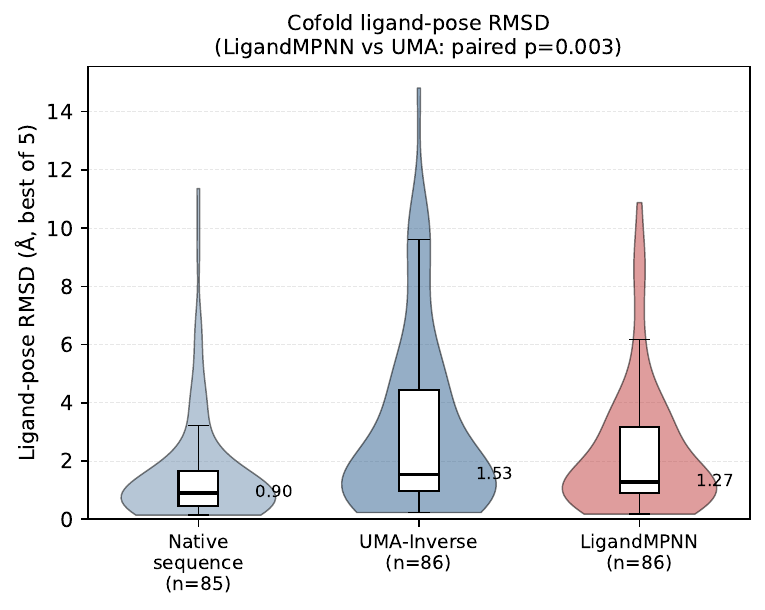}
    \caption{Boltz-2 cofold: ligand-pose RMSD vs.\ native and LigandMPNN.}
  \end{subfigure}
  \caption{\textbf{Constrained (pocket-fixed) design.} (a)~With the native pocket held fixed, both
    methods redesign the distal positions; LigandMPNN recovers them somewhat better than UMA-Inverse
    ($N{=}104$). (b)~Boltz-2 cofold ligand-pose RMSD for the native sequence, UMA-Inverse, and
    LigandMPNN: both design methods sit near the native floor and are confidently folded (ligand
    ipTM $\approx$$0.96$), with LigandMPNN slightly closer to the native pose (paired Wilcoxon
    $p{=}0.003$, $N{=}86$). Boltz-2 ipTM is a \emph{predicted} confidence, not a measurement; it is
    used here only for relative comparison between methods on the same scaffolds.}
  \label{fig:pocket}
\end{figure}

This setting mirrors a common real design task: rather than redesigning the catalytic or binding
pocket, the known pocket is held fixed, preserving its established activity, while the rest of the
protein is redesigned (Fig.~\ref{fig:pocket}). Over $104$ targets with the pocket fixed to native
identity, LigandMPNN recovers more of the distal (non-pocket) positions than UMA-Inverse
($53.7\%$ vs.\ $48.3\%$ overall; small molecule $53.5\%$ vs.\ $48.8\%$, metal $55.5\%$ vs.\ $43.7\%$),
mirroring the interface-recovery gap. At a matched sampling temperature ($T{=}0.1$) UMA-Inverse's
designs are also less diverse (mean pairwise Hamming $0.14$ vs.\ $0.19$); we report this only with the
caveat that diversity is temperature-tunable and a single nominal temperature is not directly
comparable across models, so it is not a controlled comparison.

Because native-sequence recovery penalizes any valid non-native solution (it assumes the crystal
sequence is the unique optimum, which it is not) and additionally captures non-structural signal such
as phylogeny and dataset sampling bias~\citep{shuai2025caliby}, we also validate the designs
structurally, an orthogonal check that makes no such assumption. We cofold each design with Boltz-2
\citep{passaro2025boltz2} and compare, under the identical protocol, against the native crystal
sequence (an upper bound on agreement) and LigandMPNN. Across $104$ targets (ligand pose measurable on
$86$ after heavy-atom matching), UMA-Inverse redesigns are confidently folded and
ligand-binding-competent: ligand ipTM $\approx$$0.96$ and a median best-of-5 ligand-pose RMSD of
$1.53$\,{\AA}, within $\sim$$0.6$\,{\AA} of the native-sequence floor ($0.90$\,{\AA}). They do not,
however, match LigandMPNN's geometry: LigandMPNN is modestly but significantly better on ligand pose
($1.27$\,{\AA} median; paired $\Delta{=}{+}0.13$\,{\AA}, Wilcoxon $p{=}0.003$, $N{=}86$), and likewise
on pocket-C$\alpha$ and whole-scaffold RMSD. The cofold thus confirms UMA-Inverse produces foldable,
ligand-competent redesigns but does not overturn LigandMPNN's accuracy advantage, a residual gap for a
single-pass model.

\subsection{Distogram representation and ligand-distal signal}
\label{sec:results_distal}

\begin{figure}[t]
  \centering
  \includegraphics[width=\linewidth]{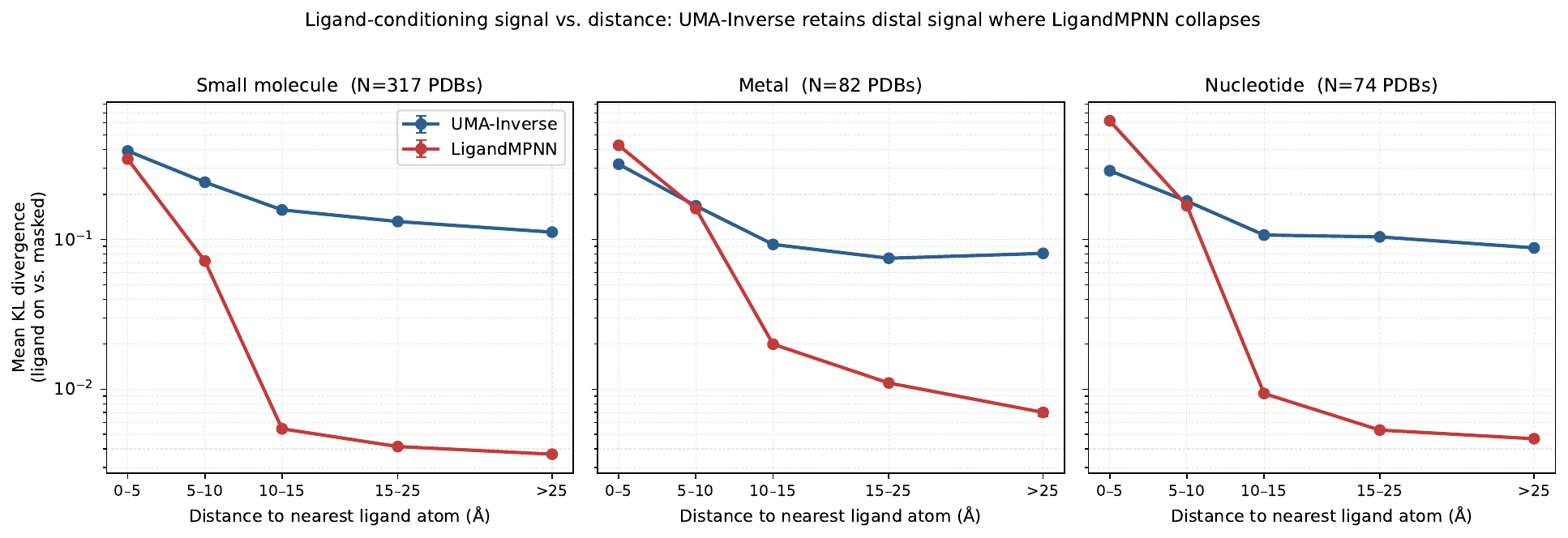}
  \caption{\textbf{Ligand-conditioning signal vs.\ distance, by ligand class.} Mean KL divergence
    between the ligand-conditioned and ligand-masked per-residue distributions in 5\,{\AA} shells
    (log scale, $\pm$1 SEM), for UMA-Inverse and LigandMPNN on the small-molecule ($N{=}317$), metal
    ($N{=}82$), and nucleotide ($N{=}74$) test splits. Both respond at the immediate interface, but
    beyond $\sim$$10$\,{\AA} LigandMPNN's signal collapses toward zero while UMA-Inverse retains a
    substantial signal: the dense pair encoder propagates ligand identity to distal residues.}
  \label{fig:distal}
\end{figure}

The auxiliary distogram head recovers binned inter-residue distances at 93.1\% top-1 accuracy,
confirming the encoder maintains an explicit geometric representation. To probe how far ligand
information reaches, we measure, in 5\,{\AA} distance shells, the KL divergence between each model's
per-residue amino-acid distribution with the ligand present versus masked (Fig.~\ref{fig:distal}).
At the immediate interface ($0$--$5$\,{\AA}) both models are strongly ligand-sensitive; LigandMPNN is
in fact the more sensitive on the metal and nucleotide splits. The two diverge sharply with distance:
beyond $10$\,{\AA} LigandMPNN's signal falls to near zero (mean KL $\le 0.02$ at $10$--$15$\,{\AA} for
all three classes), whereas UMA-Inverse retains a substantial signal ($0.09$--$0.16$ at that shell)
that persists past $25$\,{\AA}; at distal shells it stands several-fold, up to $\sim$$30\times$, above LigandMPNN's.
This \emph{distal persistence} is a representational property of the dense, all-pairs PairMixer
encoder: ligand identity is made available throughout the fold rather than confined to the binding
pocket as a local message-passing graph confines it. It should not be over-read, since the signal
does \emph{not} translate into a design advantage: LigandMPNN is more accurate both at the interface
(\S\ref{sec:results_iface}) and on pocket-fixed distal positions (\S\ref{sec:results_pocket}). What it
shows is a concrete, measurable difference in \emph{what the architecture encodes}. Whether global
ligand propagation helps would have to be tested on tasks where distal context matters beyond
single-position native recovery, such as allosteric or multi-site design, or function- rather than
identity-based objectives, which we leave to future work.

\section{Discussion}
\label{sec:discussion}

UMA-Inverse shows that a compact ($\sim$$3.3$M-parameter), distogram-supervised dense pair
encoder, with no triangle attention or sequence track, is a viable architecture for
ligand-conditioned inverse folding: it reaches within $3.7$--$9.3$\,pp of
LigandMPNN on the small-molecule and metal splits and produces foldable,
ligand-competent sequences. It does not, however, match LigandMPNN, which is more accurate on
interface recovery, on pocket-fixed distal recovery, and on cofolded ligand pose, and its
nucleic-acid routing fails to improve on the ligand-agnostic baseline for an
architectural reason we examine below. We therefore present
UMA-Inverse as a compact baseline and as the vehicle for one specific, reproducible finding, namely how a
dense all-pairs encoder distributes ligand information through a fold, rather than as a
state-of-the-art design tool. As noted in \S\ref{sec:results_pocket}, native-sequence recovery also
penalizes equally valid alternatives, so it understates any
method whose designs are foldable but non-native; our cofold analysis is the orthogonal check, and by
it both methods yield valid complexes (LigandMPNN slightly closer to the native pose).

\paragraph{What the auxiliary distogram and ligand attention buy.}
The two architectural additions target distinct failure modes. The distogram objective constrains the
encoder to keep its pair tensor structure-predictive (93.1\% top-1), rather than collapsing
it into a sequence-only shortcut; the learned ligand attention then reads that tensor in a
position-specific way, repairing the distal-residue divergence of a uniform mean-pool. These additions
are motivated by design rationale and the auxiliary head's high top-1 binned-distance accuracy rather
than a controlled ablation, a limitation we return to below.

\paragraph{Remaining gaps.}
Against our matched-protocol LigandMPNN, the metal and small-molecule gaps are modest under this
controlled comparison ($55.1\%$ vs.\ $64.4\%$ and $56.1\%$ vs.\ $59.8\%$) and smaller than LigandMPNN's
published numbers imply. The nucleotide split is the clear failure: at $35.3\%$ UMA-Inverse only matches
the ligand-agnostic ProteinMPNN ($34.0\%$), so its nucleic-acid conditioning adds essentially
nothing, even as LigandMPNN extracts a $16.5$\,pp gain from the same atoms.
An atom-context representation is not inherently at fault, since LigandMPNN also conditions on
nucleic-acid atoms and still outperforms ProteinMPNN here; the difference is architectural.

UMA-Inverse admits
ligand atoms as nodes in a single dense, all-pairs tensor, sharing one global token axis with the
residues, and because triangle multiplication scales as $O(L^3)$ in the token count $L$ the atom budget
must be small and \emph{global} (we keep the $\leq 50$ atoms nearest the protein centroid, shared by
every residue). For a compact cofactor that is the whole ligand; for a $400$--$1500$-atom nucleic acid
it is a thin, often non-interfacial slice, identical for all residues, whereas a graph model gives
each residue its own local atom neighborhood and so scales to a macromolecule. Empirically, across the
nucleotide test set the nucleic acid has a median of $526$ heavy atoms (range $57$--$1546$), so the
global $50$-atom budget retains a median of just $9\%$ of it and captures only a median $24\%$ of the
true protein--nucleic-acid interface (NA heavy atoms within $5$\,{\AA} of any protein atom). This exposes an
intrinsic tension: the dense all-pairs tensor is exactly what propagates ligand identity to distal
residues (\S\ref{sec:results_distal}) \emph{and} what makes a macromolecular ligand expensive to
represent. The same choice is both the strength and the constraint. \textbf{The most promising path is
therefore a hybrid architecture}: keep the dense residue--residue tensor for geometry and distal
propagation, but condition on ligands through a separate per-residue atom-context module injected into
the residue node features (local in that atoms bypass the cubic trunk, not in atom coverage: each
residue can still read the full ligand at its true relative geometry, unlike a $k$-NN graph's
distance-truncated neighborhood), decoupling ligand coverage from the cubic token budget. The cubic cost is not a law of nature forcing the global cap: it arises only because atoms
currently share the residue token axis, entering the triangle multiplication as apex indices and
forming atom$\times$atom pairs that inverse folding never uses. If atoms instead enter only as
residue--atom pairs (never atom$\times$atom, never as the apex index of the triangle update), the cost
falls to roughly $O(R^3 + R^2 A)$ in $R$ residues and $A$ atoms, linear in the atom count, and the
global $50$-atom budget disappears entirely, because atoms then ride a cheap linear path rather than
the cubic trunk.

\paragraph{Limitations.}
The main limitation of this evaluation is that we do not ablate the two architectural additions: we do
not isolate the contribution of the distogram head or the position-specific ligand attention to
recovery, and a controlled on/off ablation of the two is the obvious next experiment. For the same
reason we do not run the controlled test that would confirm the global atom budget is the dominant
cause of the nucleotide failure, and we do not attribute the residual small-molecule and metal gaps to
a specific architectural cause. Finally, the diversity comparison (\S\ref{sec:results_pocket}) is a
single-temperature snapshot rather than the temperature-matched Pareto comparison it would take to
settle that point.

\paragraph{Split integrity.}
Two further reproducibility issues affect the inherited splits themselves. First, several PDB accession
codes from the published LigandMPNN split could not be retrieved from RCSB via the current data API
despite correct query construction, indicating that a benchmark defined as a static list of accession
codes is not fully stable to retrieve over time, independent of any errors in our own fetching code.
Second, at least some structures in the metal test split contain metal ions with no established
biological function: 1F35 and 1JOB (murine olfactory marker protein, in two spacegroups) bind
Zn\textsuperscript{2+} that is directly attributable to the crystallization additive (200~mM zinc
acetate; \citet{smith2002omp}) rather than any native coordination site, and the primary citation
itself notes the protein surface lacks an obvious small-molecule or metal-binding site. Because
recovering the sequence context around an adventitious crystallization ion is a different task from
recovering genuine coordination chemistry, contaminated entries like these make interface recovery on
the metal split an imperfect proxy for a model's ability to design functional metal-binding sites; the
prevalence of this issue across the 82 test structures is unquantified. It is, however, a recognized
and quantifiable problem: \citet{laveglia2022metalsites} report that many metal ions in PDB structures
are adventitious rather than physiological and train a classifier that distinguishes the two at
${\sim}90\%$ accuracy for zinc. Screening train/val/test curation for crystallization-derived versus
biologically functional metal ions along these lines would sharpen this benchmark for future models.

\paragraph{Future directions.}
The most direct follow-up to our results is to test whether the encoder's distal ligand propagation
confers any practical advantage on tasks where distal context plausibly matters, such as allosteric or
multi-site design, or function- rather than identity-based objectives, since it demonstrably does not
help single-position native recovery. A controlled, temperature-matched recovery-vs-diversity (Pareto)
comparison against LigandMPNN would also place the diversity question on firmer footing than the
single-temperature snapshot reported here. Realizing the \emph{hybrid encoder} argued for above, a dense residue tensor for geometry plus a per-residue atom-context
module for ligand conditioning (full ligand coverage, off the cubic trunk), is the most direct route to fixing nucleotide design; nucleotide-level
tokens or sparse atom attention are alternatives. Other
extensions include flow-matching~\citep{lipman2022flowmatching} in place of cross-entropy, full RDKit-derived
ligand chemistry and bond topology (scaffolded but disabled in this model), and adding triangle
self-attention to the pair encoder. On the data side, the training signal itself could be improved:
better-stratified train/test splits, a larger and more recent PDB training set, and distillation from
\emph{predicted} structures (using AlphaFold/Boltz-predicted protein--ligand complexes as additional
supervision) would all expand coverage, particularly for the data-sparse nucleotide regime.

\section{Conclusion}

We introduced UMA-Inverse, a ligand- and nucleic-acid-conditioned inverse-folding model that replaces
sparse GNN message-passing with a distogram-supervised dense pair encoder and a learned
ligand-attention decoder. Trained with a three-stage curriculum on the public LigandMPNN data split, the model
achieves 66.1\% teacher-forced recovery and 56.1\%/55.1\%/35.3\%
interface recovery on the small-molecule/metal/nucleotide splits, with well-folded,
ligand-competent cofolds. It does not surpass LigandMPNN on recovery or cofolded pose, and its
nucleic-acid routing fails to improve on the ligand-agnostic baseline (extracting none of the gain LigandMPNN obtains from the same atoms), which we trace to its
dense encoder's global ligand-atom budget and argue points toward a hybrid design (\S\ref{sec:discussion}).
Its contribution is to show that a compact dense pair encoder, supervised by an auxiliary
distogram objective, is a viable alternative architecture for small-molecule and
metal partners, and to characterize its distinctive propagation of ligand information to residues far
beyond the binding pocket.

\paragraph{Code and data availability.}
Code, training and evaluation scripts, and the trained checkpoint are released at
\href{https://github.com/WSobo/UMA-Inverse}{\nolinkurl{github.com/WSobo/UMA-Inverse}}; the weights are
hosted on the Hugging Face Hub
(\href{https://huggingface.co/WSobo/UMA-Inverse}{\nolinkurl{huggingface.co/WSobo/UMA-Inverse}}) and are
auto-fetched on first use. An interactive demo (a CPU-hosted web UI and REST API serving the model)
is available at
\href{https://huggingface.co/spaces/WSobo/uma-inverse}{\nolinkurl{huggingface.co/spaces/WSobo/uma-inverse}}.
The LigandMPNN training split is available at
\href{https://github.com/dauparas/LigandMPNN}{\nolinkurl{github.com/dauparas/LigandMPNN}}.

\paragraph{Acknowledgments.}
Training compute was provided by the Yeh Laboratory at UC Santa Cruz.

\paragraph{Competing interests.}
The author declares no competing interests.

\appendix
\clearpage

\section{Per-amino-acid recovery}
\label{sec:aa_recovery}

Per-amino-acid teacher-forced recovery on the full validation set (875 PDBs, 525{,}224 residues),
with native and predicted marginal frequencies for reference (Table~\ref{tab:peraacov}). Recovery is
highest for backbone-constrained residues (Pro 0.953, Gly 0.906) and lowest for low-frequency,
context-dependent residues (Gln 0.497, Met 0.500); predicted and native compositions track closely.

\begin{table}[!ht]
\centering
\caption{Per-amino-acid teacher-forced recovery, native frequency, and predicted frequency on the
  full validation set. UMA-Inverse, best checkpoint (epoch 11).}
\label{tab:peraacov}
\begin{tabular}{lccc@{\hskip 2em}lccc}
\toprule
AA & Recovery & Native & Pred. & AA & Recovery & Native & Pred. \\
\midrule
A & 0.737 & 0.084 & 0.083 & M & 0.500 & 0.020 & 0.012 \\
C & 0.610 & 0.012 & 0.009 & N & 0.606 & 0.042 & 0.036 \\
D & 0.716 & 0.059 & 0.065 & P & 0.953 & 0.047 & 0.061 \\
E & 0.657 & 0.067 & 0.074 & Q & 0.497 & 0.036 & 0.022 \\
F & 0.687 & 0.041 & 0.037 & R & 0.609 & 0.051 & 0.047 \\
G & 0.906 & 0.072 & 0.080 & S & 0.651 & 0.059 & 0.062 \\
H & 0.553 & 0.024 & 0.016 & T & 0.673 & 0.053 & 0.057 \\
I & 0.693 & 0.057 & 0.054 & V & 0.769 & 0.072 & 0.079 \\
K & 0.603 & 0.056 & 0.055 & W & 0.592 & 0.015 & 0.011 \\
L & 0.814 & 0.097 & 0.107 & Y & 0.645 & 0.035 & 0.032 \\
\bottomrule
\end{tabular}
\end{table}

\section{Validation metrics by training stage}
\label{sec:training_detail}

\begin{table}[!ht]
\centering
\caption{Final validation metrics at the end of each curriculum stage.}
\label{tab:training_stages}
\begin{tabular}{lccccc}
\toprule
Stage & GPUs & Max nodes & Best epoch & Val accuracy & Val loss \\
\midrule
Stage 1 & 1$\times$A5500 & 64  & 4  & 40.5\% & 2.034 \\
Stage 2 & 2$\times$A100  & 128 & 12 & 51.3\% & 1.635 \\
Stage 3 & 2$\times$A100  & 384 & 11 & 63.2\% & 1.217 \\
\bottomrule
\end{tabular}
\end{table}

\clearpage
\bibliographystyle{unsrtnat}
\bibliography{references}

\begin{thebibliography}{21}
\providecommand{\natexlab}[1]{#1}
\providecommand{\url}[1]{\texttt{#1}}
\expandafter\ifx\csname urlstyle\endcsname\relax
  \providecommand{\doi}[1]{doi: #1}\else
  \providecommand{\doi}{doi: \begingroup \urlstyle{rm}\Url}\fi

\bibitem[Dauparas et~al.(2022)Dauparas, Anishchenko, Bennett, Bai, Ragotte,
  Milles, Wicky, Courbet, de~Haas, Bethel, Leung, Huddy, Pellock, Tischer,
  Chan, Gross, Bhatt, Kang, et~al.]{dauparas2022proteinmpnn}
Justas Dauparas, Ivan Anishchenko, Nathaniel Bennett, Hua Bai, Robert~J.
  Ragotte, Lukas~F. Milles, Basile I.~M. Wicky, Alexis Courbet, Rob~J. de~Haas,
  Neville Bethel, Philip J.~Y. Leung, Timothy~F. Huddy, Samuel Pellock, Doug
  Tischer, Frederick Chan, Brian Gross, Vikram Bhatt, Asim Kang, et~al.
\newblock Robust deep learning--based protein sequence design using
  {ProteinMPNN}.
\newblock \emph{Science}, 378\penalty0 (6615):\penalty0 49--56, 2022.
\newblock \doi{10.1126/science.add2187}.

\bibitem[Dauparas et~al.(2025)Dauparas, Lee, Pecoraro, An, Anishchenko,
  Glasscock, and Baker]{dauparas2025ligandmpnn}
Justas Dauparas, Gyu~Rie Lee, Robert Pecoraro, Linna An, Ivan Anishchenko,
  Cameron Glasscock, and David Baker.
\newblock Atomic context-conditioned protein sequence design using
  {LigandMPNN}.
\newblock \emph{Nature Methods}, 22\penalty0 (4):\penalty0 717--723, 2025.
\newblock \doi{10.1038/s41592-025-02626-1}.
\newblock Preprint: bioRxiv 2023.12.22.573103.

\bibitem[Yi et~al.(2025)Yi, Jamali, and Scheres]{yi2025adflip}
Kai Yi, Kiarash Jamali, and Sjors H.~W. Scheres.
\newblock All-atom inverse protein folding through discrete flow matching.
\newblock In \emph{International Conference on Machine Learning (ICML)}, 2025.
\newblock arXiv:2507.14156.

\bibitem[Jumper et~al.(2021)Jumper, Evans, Pritzel, Green, Figurnov,
  Ronneberger, Tunyasuvunakool, Bates, {\v{Z}}{\'{\i}}dek,
  et~al.]{jumper2021alphafold}
John Jumper, Richard Evans, Alexander Pritzel, Tim Green, Michael Figurnov,
  Olaf Ronneberger, Kathryn Tunyasuvunakool, Russ Bates, Augustin
  {\v{Z}}{\'{\i}}dek, et~al.
\newblock Highly accurate protein structure prediction with {AlphaFold}.
\newblock \emph{Nature}, 596\penalty0 (7873):\penalty0 583--589, 2021.
\newblock \doi{10.1038/s41586-021-03819-2}.

\bibitem[Abramson et~al.(2024)Abramson, Adler, Dunger, Evans, Green, Pritzel,
  Ronneberger, Willmore, Ballard, Bambrick, et~al.]{abramson2024alphafold3}
Josh Abramson, Jonas Adler, Jack Dunger, Richard Evans, Tim Green, Alexander
  Pritzel, Olaf Ronneberger, Lindsay Willmore, Andrew~J. Ballard, Joshua
  Bambrick, et~al.
\newblock Accurate structure prediction of biomolecular interactions with
  {AlphaFold~3}.
\newblock \emph{Nature}, 630:\penalty0 493--500, 2024.
\newblock \doi{10.1038/s41586-024-07487-w}.

\bibitem[Ouyang-Zhang et~al.(2025)Ouyang-Zhang, Murugan, Diaz, Scarpellini,
  Bowen, Gruver, Klivans, Kr{\"a}henb{\"u}hl, Faust, and
  Al-Shedivat]{ouyangzhang2025pairmixer}
Jeffrey Ouyang-Zhang, Pranav Murugan, Daniel~J. Diaz, Gianluca Scarpellini,
  Richard~Strong Bowen, Nate Gruver, Adam Klivans, Philipp Kr{\"a}henb{\"u}hl,
  Aleksandra Faust, and Maruan Al-Shedivat.
\newblock {Triangle Multiplication Is All You Need For Biomolecular Structure
  Representations}.
\newblock arXiv:2510.18870 [q-bio.QM], 2025.
\newblock Genesis Research; UT Austin.

\bibitem[Jing et~al.(2021)Jing, Eismann, Suriana, Townshend, and
  Dror]{jing2021gvp}
Bowen Jing, Stephan Eismann, Patricia Suriana, Raphael J.~L. Townshend, and Ron
  Dror.
\newblock Learning from protein structure with geometric vector perceptrons.
\newblock In \emph{International Conference on Learning Representations
  (ICLR)}, 2021.
\newblock URL \url{https://openreview.net/forum?id=1YLJDvSx6J4}.

\bibitem[Hsu et~al.(2022)Hsu, Verkuil, Liu, Lin, Hie, Sercu, Lerer, and
  Rives]{hsu2022esmif}
Chloe Hsu, Robert Verkuil, Jason Liu, Zeming Lin, Brian Hie, Tom Sercu, Adam
  Lerer, and Alexander Rives.
\newblock Learning inverse folding from millions of predicted structures.
\newblock In \emph{International Conference on Machine Learning (ICML)}, pages
  8946--8970. PMLR, 2022.

\bibitem[Gao et~al.(2023)Gao, Tan, and Li]{gao2023pifold}
Zhangyang Gao, Cheng Tan, and Stan~Z. Li.
\newblock {PiFold}: Toward effective and efficient protein inverse folding.
\newblock In \emph{International Conference on Learning Representations
  (ICLR)}, 2023.
\newblock URL \url{https://openreview.net/forum?id=oMsN9TYwJ0j}.

\bibitem[Ren et~al.(2024)Ren, Yu, Bu, and Zhang]{ren2024carbondesign}
Milong Ren, Chungong Yu, Dongbo Bu, and Haicang Zhang.
\newblock Accurate and robust protein sequence design with {CarbonDesign}.
\newblock \emph{Nature Machine Intelligence}, 6:\penalty0 536--547, 2024.
\newblock \doi{10.1038/s42256-024-00838-2}.

\bibitem[Passaro et~al.(2025)Passaro, Corso, Wohlwend, Reveiz, Thaler, Somnath,
  Getz, Portnoi, Roy, Stark, Kwabi-Addo, Beaini, Jaakkola, and
  Barzilay]{passaro2025boltz2}
Saro Passaro, Gabriele Corso, Jeremy Wohlwend, Mateo Reveiz, Stephan Thaler,
  Vignesh~Ram Somnath, Noah Getz, Tally Portnoi, Julien Roy, Hannes Stark,
  David Kwabi-Addo, Dominique Beaini, Tommi Jaakkola, and Regina Barzilay.
\newblock {Boltz-2}: Towards accurate and efficient binding affinity
  prediction.
\newblock \emph{bioRxiv}, 2025.
\newblock \doi{10.1101/2025.06.14.659707}.

\bibitem[Watson et~al.(2023)Watson, Juergens, Bennett, Trippe, Yim, Eisenach,
  Ahern, Borst, Ragotte, Milles, et~al.]{watson2023rfdiffusion}
Joseph~L. Watson, David Juergens, Nathaniel~R. Bennett, Brian~L. Trippe, Jason
  Yim, Helen~E. Eisenach, Woody Ahern, Andrew~J. Borst, Robert~J. Ragotte,
  Lukas~F. Milles, et~al.
\newblock De novo design of protein structure and function with {RFdiffusion}.
\newblock \emph{Nature}, 620:\penalty0 1089--1100, 2023.
\newblock \doi{10.1038/s41586-023-06415-8}.

\bibitem[Butcher et~al.(2025)Butcher, Krishna, Mitra, Brent, Li, Corley, Kim,
  Funk, Mathis, Salike, et~al.]{butcher2025rfdiffusion3}
J.~K.~V. Butcher, R.~Krishna, R.~Mitra, R.~I. Brent, Y.~Li, N.~Corley, P.~Kim,
  J.~Funk, S.~V. Mathis, S.~Salike, et~al.
\newblock De novo design of all-atom biomolecular interactions with
  {RFdiffusion3}.
\newblock \emph{bioRxiv}, 2025.
\newblock \doi{10.1101/2025.09.18.676967}.

\bibitem[Shuai et~al.(2025)Shuai, Lu, Bhatti, Kouba, and
  Huang]{shuai2025caliby}
Richard~W. Shuai, Tianyu Lu, Subhang Bhatti, Petr Kouba, and Po-Ssu Huang.
\newblock Ensemble-conditioned protein sequence design with {Caliby}.
\newblock \emph{bioRxiv}, 2025.
\newblock \doi{10.1101/2025.09.30.679633}.

\bibitem[Krapp et~al.(2024)Krapp, Meireles, Abriata, and
  Dal~Peraro]{krapp2024carbonara}
Lucien~F. Krapp, Fernando~A. Meireles, Luciano~A. Abriata, and Matteo
  Dal~Peraro.
\newblock Context-aware geometric deep learning for protein sequence design.
\newblock \emph{Nature Communications}, 15:\penalty0 6273, 2024.
\newblock \doi{10.1038/s41467-024-50571-y}.

\bibitem[Kubaney et~al.(2025)Kubaney, Favor, McHugh, Mitra, Pecoraro, Dauparas,
  Glasscock, and Baker]{kubaney2025nampnn}
Andrew Kubaney, Andrew Favor, Lilian McHugh, Raktim Mitra, Robert Pecoraro,
  Justas Dauparas, Cameron Glasscock, and David Baker.
\newblock {RNA} sequence design and protein--{DNA} specificity prediction with
  {NA-MPNN}.
\newblock \emph{bioRxiv}, 2025.
\newblock \doi{10.1101/2025.10.03.679414}.

\bibitem[Wei et~al.(2025)Wei, Guerrini, and Eberini]{wei2025ifbenchmark}
Yao Wei, Uliano Guerrini, and Ivano Eberini.
\newblock Benchmarking and consensus ranking of inverse folding models for
  protein--ligand interface design.
\newblock In \emph{Companion Proceedings of the 16th ACM International
  Conference on Bioinformatics, Computational Biology and Health Informatics
  (BCB Companion)}, 2025.
\newblock \doi{10.1145/3768322.3769031}.

\bibitem[Loshchilov and Hutter(2019)]{loshchilov2019adamw}
Ilya Loshchilov and Frank Hutter.
\newblock Decoupled weight decay regularization.
\newblock In \emph{International Conference on Learning Representations
  (ICLR)}, 2019.
\newblock URL \url{https://openreview.net/forum?id=Bkg6RiCqY7}.

\bibitem[Smith et~al.(2002)Smith, Firestein, and Hunt]{smith2002omp}
Paul~C. Smith, Stuart Firestein, and John~F. Hunt.
\newblock The crystal structure of the olfactory marker protein at {2.3}~{\aa}
  resolution.
\newblock \emph{Journal of Molecular Biology}, 319\penalty0 (3):\penalty0
  807--821, 2002.
\newblock \doi{10.1016/S0022-2836(02)00242-5}.

\bibitem[Laveglia et~al.(2022)Laveglia, Giachetti, Sala, Andreini, and
  Rosato]{laveglia2022metalsites}
Vincenzo Laveglia, Andrea Giachetti, Davide Sala, Claudia Andreini, and Antonio
  Rosato.
\newblock Learning to identify physiological and adventitious metal-binding
  sites in the three-dimensional structures of proteins by following the hints
  of a deep neural network.
\newblock \emph{Journal of Chemical Information and Modeling}, 62\penalty0
  (12):\penalty0 2951--2960, 2022.
\newblock \doi{10.1021/acs.jcim.2c00522}.

\bibitem[Lipman et~al.(2022)Lipman, Chen, Ben-Hamu, Nickel, and
  Le]{lipman2022flowmatching}
Yaron Lipman, Ricky T.~Q. Chen, Heli Ben-Hamu, Maximilian Nickel, and Matt Le.
\newblock Flow matching for generative modeling.
\newblock arXiv:2210.02747 [cs.LG], 2022.

\end{thebibliography}

\end{document}